\documentclass[12pt]{article}
\usepackage{graphics}
\begin{document}
\begin{center}
{Relation between the Vertex Function and the Propagators  \\
in the Pseudovector Coupling Pion-Nucleon System}
\end{center}
\begin{center}
{Susumu Kinpara}
\end{center}
\begin{center}
{\it National Institute of Radiological Sciences \\ Chiba 263-8555, Japan}
\end{center}
\begin{abstract}
The procedure for deriving the Ward-Takahashi identity is applied to the pseudovector coupling pion-nucleon system
in which the isovector and axial-vector current does not conserved.
The relation is given in closed form although it is more complicated than that of the quantum electrodynamics.
The application to the propagator has revealed the interesting results.
\end{abstract}
\section*{\normalsize{1 \quad Introduction}}
\hspace*{4.mm}
Recently we have suggested a potential model on nucleon-nucleon interaction 
derived from the Bethe-Salpeter (BS) equation $\cite{Kinpara}$.
The BS amplitude is expanded by a set of the Gamma matrices and it results in the simultaneous equations.
At the relative time $t$ = 0 they are analogous to the Schr$\ddot {\rm o}$dinger equation  
and therefore which make us apply to two-nucleon system not only for the bound state also the elastic scattering.  
\\\hspace*{4.mm}
The spin observable is one of the subjects investigating the applicability of the potential.
Particularly to determine the spin singlet part among two possible components in the equations 
we need to compare each result with the experimental data.
For example in the case of the proton-proton ($p$-$p$) elastic scattering 
the axial-vector equation is found to describe the correlation parameter better than the pseudoscalar equation.    
\\\hspace*{4.mm}
When we calculate the spin observable there is an issue for the $p$-$p$ elastic scattering.
The dependence of the pseudovector coupling constant $f_\pi$ of the pion-nucleon interaction 
on the energy of the incident nucleon becomes large such as $f_\pi \sim 1/20$ at the intermediate energy region.
Meanwhile such a drastic change of $f_\pi$ from the empirical value $f_\pi$ = 1 
is not observed for the proton-neutron ($p$-$n$) elastic scattering.
The quantum effects may cause the difference between the $p$-$p$ and the $p$-$n$ ones 
having an result on the value of the parameter like the coupling constant.
Therefore the apparent breaking of the charge independence should be understood 
by taking account of the higher-order quantum corrections.
\\\hspace*{4.mm}
The BS equation is essentially non-perturbative 
and then it is applicable to any two-body system irrespective of the strength of the interaction.
To supplement the framework with the higher-order corrections 
the non-perturbative treatment of the quantum field is also needed. 
Particularly the ladder approximation for the irreducible kernel is the lowest-order in the series of the coupling constant 
and the correction is one of the subjects for us to calculate the quantities on nucleon-nucleon system exactly.
In the present study we will investigate the propagators of the pseudovector coupling pion-nucleon system 
using the relation accompanying the non-perturbative term.
\hspace*{4.mm}
\section*{\normalsize{2 \quad The non-perturbative relation }}
\hspace*{4.mm}
The pseudovector coupling is one of the two possible types of pion-nucleon interaction to construct the vertex part.
The interacting lagrangian density is given by
\begin{eqnarray}
L(x) = -\frac{f_\pi}{m_\pi}\,\bar{\psi}(x)\,\gamma_5\,\gamma_\mu\,\tau_i\,\psi(x)\,\partial^\mu\phi_i(x),
\end{eqnarray}
in which $f_\pi$ and $m_\pi$ are the coupling constant and the mass of pion respectively.
For pion is the isovector and pseudoscalar boson the set of the three fields $\phi_i(x)$ ({\it i} = 1, 2, 3) is coupled 
with the nucleon field $\psi(x)$ through the usual isospin matrix $\tau_i$.
\\\hspace*{4.mm}
To investigate the non-perturbative property of the propagators 
we pay attention to the relation between the vertex function and the self-energy of nucleon.
In quantum electrodynamics (QED) it is known as the Ward-Takahashi identity which is indispensable to accomplish the calculation of the higher-order quantum corrections $\cite{Nishijima}$.
As will be seen in the following the extension to the pseudovector coupling pion-nucleon system is not necessarily straightforward since the quantity $J^{5 \mu}_i(x)\equiv\bar{\psi}(x)\,\gamma^\mu\,\gamma_5\,\tau_i\,\psi(x)$ 
connected to the pion field does not construct the conserved current ($\partial \cdot J^{5}_i(x) \neq 0$).
\\\hspace*{4.mm}
Under the interaction given by Eq. (1) the equation of motion for pion field becomes 
\begin{eqnarray}
(\partial^2+m_\pi^2)\phi_i(x)=-\frac{f_\pi}{m_\pi}\partial \cdot J^{5}_i(x),
\end{eqnarray}
and the equal-time commutation relations are
\begin{eqnarray}
\delta(x_0-y_0)[\,\phi_i(x),\,\frac{\partial \phi_j(y)}{\partial y_0}\,] = i\,\delta_{ij}\,\delta^4(x-y).\qquad(i,j=1,2,3)
\end{eqnarray}
The equal-time commutation relations between the zeroth component $J^{5\,0}_i(x)$ 
and the Dirac fields $\varphi_a(x_a)$ ($\,\equiv \, \psi(x_a)$ or $\bar{\psi}^{\small T}(x_a)\,$) are given 
\begin{eqnarray}
\delta(x_{a0}-x_0)[\,\varphi_a(x_a),\,J^{5\,0}_i(x)\,] = e^i_{a}\,\varphi_a(x_a)\delta^4(x_a-x),
\end{eqnarray}
in which $e^i_a = \gamma_5 \tau_i$ ($\varphi_a(x_a)=\psi(x_a)$) 
or $e^i_a = \gamma_5 \tau_i^{\small T}$ ($\varphi_a(x_a)=\bar{\psi}^{\small T}(x_a)$).
\\\hspace*{4.mm}
Our interest is to derive the generalized relation including any number of operators
in the Heisenberg representation analogous to that of QED $\cite{Nishijima}$.
The quantity of the T-product constructed by the operators is introduced as
\begin{eqnarray}
T[\phi_i(x)\cdots\varphi_a(x_a)\cdots].
\end{eqnarray}
By operating $\partial^2+m_\pi^2$ on the quantity (5) and using Eqs. (2) and (3) it results in 
\begin{eqnarray}
&&(\partial^2+m_\pi^2)\,T[\phi_i(x)\cdots\varphi_a(x_a)\cdots]   \qquad\nonumber\\          
&&\qquad= -\frac{f_\pi}{m_\pi}\,T[\partial \cdot J^{5}_i(x) \cdots]-i\frac{\delta}{\delta\,\phi_i(x)}\,T[\cdots],
\end{eqnarray}
with the definition of the functional derivative
\begin{eqnarray}
\frac{\delta\,\phi_j(x^\prime)}{\delta\,\phi_i(x)} = \delta_{ij}\,\delta^4(x-x^\prime).
\end{eqnarray}
For Eq. (6) the relation about the divergence on $T[J^{5}_i(x)\cdots]$ shown below
\begin{eqnarray}
\partial\cdot T[J^{5}_i(x)\cdots] 
= T[\partial \cdot J^{5}_i(x)\cdots] \,-\sum_{a,b,\cdots}\,e^i_a \,\delta^4(x-x_a)\,T[\cdots],
\end{eqnarray}
is used and thus the generalized relation for the pseudovector coupling pion-nucleon system yields
\begin{eqnarray}
(\partial^2+m_\pi^2)\,T[\phi_i(x)\cdots\varphi_a(x_a)\cdots]             
= -\frac{f_\pi}{m_\pi}\partial \cdot T[J^{5}_i(x) \cdots]-i\frac{\delta}{\delta\,\phi_i(x)}T[\cdots]\nonumber\\
-\frac{f_\pi}{m_\pi}\sum_{a,b,\cdots}\,e^i_a \,\delta^4(x-x_a)\,T[\cdots].\qquad
\end{eqnarray}
\hspace*{4.mm}
In the case of the T-product constructed by two operators of the field of pion $T[\,\phi_i(x)\,\phi_j(y)\,]$ 
the relation Eq. (9) results in
\begin{eqnarray}
(\partial^2+m_\pi^2)\,\Delta_{ij}(x-y) 
= i \,\frac{f_\pi}{m_\pi}\, \partial\cdot\langle\,T[J^{5}_i(x)\,\phi_j(y)]\,\rangle -\delta_{ij}\,\delta^4(x-y),
\end{eqnarray}
\begin{eqnarray}
i\,\Delta_{ij}(x-y)\,\equiv\,\langle\,T[\,\phi_i(x)\,\phi_j(y)\,]\,\rangle,
\end{eqnarray}
which represents the equation of motion for the exact propagator $\Delta_{ij}(x-y)$.
Here the expectation value of the operator for instance $O$ in the Heisenberg representation is defined by 
$\langle \,O\, \rangle \equiv \, <0\,\vert \,O\, \vert \,0>/<0\,\vert\,0>$ using the vacuum state $\vert\,0>$.
\\\hspace*{4.mm}
When we apply the generalized relation in Eq. (9) 
to the case of the combination of one pion and two nucleons that is $T[\,\phi_i(z)\,\psi(x)\,\bar{\psi}(y)\,]$
an interesting result is obtained about the pion-nucleon vertex.
Following the procedure for QED the expectation value 
is expressed by means of the vertex function $\Gamma_i(x\,y\,;z)$ as
\begin{eqnarray}
&&\langle\,T[\,\phi_i(z)\,\psi(x)\,\bar{\psi}(y)\,]\,\rangle \\
&&=-\frac{f_\pi}{m_\pi}\int d^4 x^\prime d^4 y^\prime d^4 z^\prime
G(x-x^\prime)\Gamma_j(x^\prime\,y^\prime\,;z^\prime)G(y^\prime-y)\Delta_{ji}(z^\prime-z), \nonumber
\end{eqnarray}
with the exact pion propagator $\Delta_{ji}(z^\prime-z)$ and the exact nucleon propagator $G(x-y)$ given by
\begin{eqnarray}
i\,G(x-y) \equiv \langle\, T[\,\psi(x)\,\bar{\psi}(y)\,] \,\rangle.
\end{eqnarray}
Substituting Eq. (12) into Eq. (9) and using Eq. (10) the relation between the vertex function and the propagators
is obtained 
\begin{eqnarray}
&&\int d^4 x^\prime d^4 y^\prime\,
G(x-x^\prime)\,\Gamma_i(x^\prime\,y^\prime\,;z)\,G(y^\prime-y)
+\partial_z\cdot\langle\, T[\,\psi(x)\,\bar{\psi}(y)\,J^{5}_i(z)\,]\,\rangle \nonumber\\
&&-i\frac{f_\pi}{m_\pi}\partial_z \cdot \int d^4 x^\prime d^4 y^\prime d^4 z^\prime
G(x-x^\prime)\Gamma_j(x^\prime\,y^\prime\,;z^\prime)G(y^\prime-y)
\langle T[\,\phi_j(z^\prime)\,J^{5}_i(z)\,]\rangle \nonumber\\
&&=-i\,\delta^4(z-y)\,G(x-y)\,\gamma_5\,\tau_i-i\,\delta^4(z-x)\,\tau_i\,\gamma_5\,G(x-y).
\end{eqnarray}
As shown by the techniques of the Feynman diagram 
the third term in the left side is the composite diagram of the vertex function and the polarization function for the pion propagator which are connected by one pion propagator.
Then it cancels out the same part in the second term.  
\\\hspace*{4.mm}
The right side of Eq. (14) arises from the definition of the T-product and therefore independent of the perturbative
treatment.
When we calculate the vertex function the terms play a role in it along with the perturbative part of the proper vertex.
The non-perturbative effect is interesting since by which the result of the higher-order corrections are changed.
Eq. (14) is converted to the expression in momentum space 
\begin{eqnarray}
\Gamma_i(p,q) = \Gamma_i(p,q)_{per} + \Gamma_i(p,q)_n,
\end{eqnarray}
\begin{eqnarray}
\Gamma_i(p,q)_n \equiv -i\,G(p)^{-1}\,\tau_i\,\gamma_5 -i\,\tau_i\,\gamma_5\,G(q)^{-1},
\end{eqnarray}
where $\Gamma_i(p,q)$ and $G(p)$ are the Fourier transform of $\Gamma_i(x,y\,;z)$ and $G(x)$ respectively.
$\Gamma_i(p,q)_{per}$ is the proper vertex function in momentum space defined by the perturbative expansion in the series of 
$f_\pi/m_\pi$.
The zeroth order is $\Gamma_i(p,q)_{per} = -i\,\tau_i\,\gamma_5\,\gamma\cdot(p-q) + O((f_\pi/m_\pi)^2)$
giving the usual vertex part of the pseudovector coupling interaction.
\\\hspace*{4.mm}
Restricted to the on-shell of the four-momenta $p$ and $q$ the non-perturbative part $\Gamma_i(p,q)_n$ does not contribute
to the transition element $\bar{u}(p) \Gamma_i(p,q)_n u(q)$ between the incoming and outgoing nucleons 
and then also to the Born term of the elastic scattering.  
It is not necessarily right concerning the higher-order process as will be examined in the next section. 
\section*{\normalsize{3 \quad The non-perturbative effects on the propagators}}
\section*{\small{3.1 \quad In the case of the polarization function of pion}}
\hspace*{4.mm}
The exact pion propagator $\Delta(q)$ is determined by the Dyson equation in momentum space as  
\begin{eqnarray}
\Delta(q) = \Delta^{0}(q) + \Delta^{0}(q)\,\Pi(q)\,\Delta(q).
\end{eqnarray}
The solution is given by
\begin{eqnarray}
\Delta(q) = \frac{1}{\Delta^{0}(q)^{-1}-\Pi(q)},
\end{eqnarray}
\begin{eqnarray}
\Delta^{0}(q) = \frac{1}{q^2 - m_\pi^2 + i\,\epsilon},
\end{eqnarray}
with the free pion propagator $\Delta^{0}(q)$ and the polarization function $\Pi(q)$
which are $3\times3$ matrices with respect to the isospin $\cite{Serot}$.
\\\hspace*{4.mm}
The $i\,j$ component of the exact polarization function $\Pi(q)$ is found to be
\begin{eqnarray}
\Pi_{ij}(q) = (\frac{f_\pi}{m_\pi})^2\int\frac{d^4 k}{(2 \pi)^4}
{\rm Tr}[\,\tau_i\,\gamma_5\gamma\cdot q\,G(q+k)\,\Gamma_j(q+k,k)\,G(k)\,],
\end{eqnarray}
using the vertex function $\Gamma_i(p,q)$ in Eq. (15). 
$\Pi(q)$ is divided into two parts $\Pi(q) = \Pi(q)_{per}+\Pi(q)_{n}$
where $\Pi(q)_{per}$ is the one for which ${\Gamma_j}(q+k,k)_{per}$ is used in place of $\Gamma_j(q+k,k)$ in Eq. (20).
$\Pi(q)_{n}$ is the remaining part which becomes zero unless the non-pertubative part ${\Gamma_j}(p,q)_{n}$ exists.
It is shown that by using the form of ${\Gamma_j}(p,q)_{n}$ 
the non-perturbative part $\Pi(q)_{n}$ vanishes.
Then the perturbative expansion is enough to obtain the result of $\Pi(q)$.
\\\hspace*{4.mm}
In order to calculate the polarization function $\Pi^{0}(q)_{per}$ by the lowest-order approximation 
and to obtain the convergent result the dimensional regularization method is used.
The integral on $k$ is convergent by shifting the dimension of the momentum space 
as $4 \rightarrow n \equiv 4-\epsilon$ with the infinitesimal quantity $\epsilon$.
Also the mass parameter $\mu$ is introduced to keep the mass dimension of the lagrangian density 
at $n$, so the coupling constant $f_\pi$ is multiplied by $\mu^{\epsilon/2}$.
The result of $\Pi^{0}_{ij}(q)_{per} = \delta_{ij}\,\Pi^{0}(q)_{per}$ is given by
\begin{eqnarray}
\Pi^{0}(q)_{per} = \frac{1}{\pi^2}(\frac{f_\pi}{m_\pi})^2\,M^2\,q^2\,(-\frac{2}{\epsilon}+I(q)),
\end{eqnarray}
\begin{eqnarray}
I(q) \equiv \int_0^1 dz \, {\rm log} \, \frac{M^2-z(1-z)q^2}{4 \pi \mu^2},
\end{eqnarray}
where $\it M$ is the mass of nucleon.
The explicit form of $I(q)$ is shown in Appendix 1.
\\\hspace*{4.mm}
The renormalized polarization function $\Pi^{R}(q)$ is determined so as to satisfy the renormalization conditions
at the on-shell ($q^2\,$=$\,m_\pi^2$) for the four-momentum $q$ of pion
\begin{eqnarray}
\Pi^{R}(q)\,\vert_{q^2=m_\pi^2} = 0,
\end{eqnarray}
\begin{eqnarray}
\frac{\partial\Pi^{R}(q)}{\partial q^2}\,\vert_{q^2=m_\pi^2} = 0,
\end{eqnarray}
by adding the counter terms for the mass and the field of pion to the lagrangian density.
Thus the final result becomes 
\begin{eqnarray}
&&\Pi^{R}(q) = -\frac{1}{\pi^2}\,(\frac{f_\pi}{m_\pi})^2\,M^2\,(I_0+m_\pi^2\,I_1)\,(q^2-m_\pi^2)\nonumber\\
&&+\frac{1}{\pi^2}\,(\frac{f_\pi}{m_\pi})^2\,M^2\,(q^2\,I(q)-m_\pi^2\,I_0),
\end{eqnarray}
where $I_0$ and $I_1$ are the first and the second coefficients of $I(q)$ 
expanded in the series of $q^2-m_\pi^2$ as $I(q) = I_0 +I_1\,(q^2-m_\pi^2)+O((q^2-m_\pi^2)^2)$.
The values of the coefficients $I_0$ and $I_1$ are given in Appendix 1.
\section*{\small{3.2 \quad In the case of the self-energy of nucleon }}
\hspace*{4.mm}
The nucleon propagator in momentum space is represented by the following equation
\begin{eqnarray}
G(p) = G^{0}(p) + G^{0}(p)\,\Sigma(p)\,G(p),
\end{eqnarray}
and then the exact propagator is given by
\begin{eqnarray}
G(p) = \frac{1}{G^{0}(p)^{-1}-\Sigma(p)},
\end{eqnarray}
using the free propagator
\begin{eqnarray}
G^{0}(p) = \frac{1}{\gamma\cdot p - M + i\,\epsilon}.
\end{eqnarray}
\hspace*{4.mm}
The self-energy $\Sigma(p)$ is important to construct the quantum system having the pion-nucleon interaction.
By operating $i\,\gamma\cdot\partial-M$ on $G(x-y)$ 
and using the equation of motion for the nucleon field $\psi(x)$ in conjunction with Eq. (12) 
the counterpart of Eq. (26) gives the form of $\Sigma(p)$ in momentum space as
\begin{eqnarray}
\Sigma(p) = (\frac{f_\pi}{m_\pi})^2\,\tau_i\,\gamma_5\,\gamma\cdot\int\frac{d^4 k}{(2 \pi)^4}\, k\,\Delta_{ij}(k)
\,G(p-k)\,\Gamma_j(p-k,p).
\end{eqnarray}
Owing to Eqs. (15) and (16) it is divided into three parts 
\begin{eqnarray}
\Sigma(p) = \Sigma(p)_1 + \Sigma(p)_{21} + \Sigma(p)_{22},
\end{eqnarray}
\begin{eqnarray}
\Sigma(p)_1 \equiv (\frac{f_\pi}{m_\pi})^2 \,\tau_i\,\gamma_5\,\gamma\cdot \int\frac{d^4 k}{(2 \pi)^4}\, k\,\Delta_{ij}(k)
\,G(p-k)\,\Gamma_j(p-k,p)_{per},
\end{eqnarray}
\begin{eqnarray}
\Sigma(p)_{21} \equiv -i\,(\frac{f_\pi}{m_\pi})^2 \,\tau_i\,\tau_j\,\gamma_5\,\gamma\,\gamma_5\cdot
\int\frac{d^4 k}{(2 \pi)^4} \,k\,\Delta_{ij}(k),
\end{eqnarray}
\begin{eqnarray}
\Sigma(p)_{22} \equiv -i\,(\frac{f_\pi}{m_\pi})^2 \,\tau_i\,\tau_j\,\gamma_5\,\gamma\cdot 
\int\frac{d^4 k}{(2 \pi)^4}\,k\,\Delta_{ij}(k)\,G(p-k)\,\gamma_5\,G(p)^{-1}.
\end{eqnarray}
In the present study $\Gamma_j(p-k,p)_{per}$ is assumed to represent the sum of the perturbative expansion of the usual vertex function and then $\Sigma(p)_1$ is calculated in order using the method of the Feynman diagram.
Since the integrand of $\Sigma(p)_{21}$ is anti-symmetric under the replacement $k\rightarrow-k$ 
it does not contribute to the integral and results in $\Sigma(p)_{21}=0$.
\\\hspace*{4.mm}
Besides $\Sigma(p)_1$ which is stemmed from the perturbative part the self-energy $\Sigma(p)$ contains 
the non-perturbative part $\Sigma(p)_{22}$. 
By virtue of the inverse of the nucleon propagator $G(p)^{-1}=\gamma \cdot p -M-\Sigma(p)$ in $\Sigma(p)_{22}$ 
the expression for $\Sigma(p)$ is transformed as
\begin{eqnarray}
\Sigma(p) = \frac{A(p)}{B(p)},
\end{eqnarray}
\begin{eqnarray}
A(p) \equiv i(\frac{f_\pi}{m_\pi})^2 \,\tau_i\,\gamma_5\,\gamma\cdot \int\frac{d^4 k}{(2 \pi)^4}\, k\,\Delta_{ij}(k)
\,G(p-k)\qquad\\\nonumber
(-i\,\Gamma_j(p-k,p)_{per}-\,\tau_j\,\gamma_5\,(\gamma \cdot p-M)),
\end{eqnarray}
\begin{eqnarray}
B(p) \equiv 1 - i(\frac{f_\pi}{m_\pi})^2 \,\tau_i\,\gamma_5\,\gamma\cdot 
\int\frac{d^4 k}{(2 \pi)^4}\,k\,\Delta_{ij}(k)\,G(p-k)\,\tau_j\,\gamma_5.
\end{eqnarray}
\\\hspace*{4.mm}
In order to determine the form of $\Sigma(p)$ explicitly
the approximate propagators and the vertex part are substituted for the exact ones as 
\begin{eqnarray}
\Delta_{ij}(k)\rightarrow\delta_{ij}\,\Delta^{0}(k),
\end{eqnarray}
\begin{eqnarray}
G(p-k)\rightarrow G^{0}(p-k),
\end{eqnarray}
\begin{eqnarray}
\Gamma_j(p-k,p)_{per}\rightarrow \,i\,\tau_j\,\gamma_5\gamma\cdot k.
\end{eqnarray}
Under the replacement of Eqs. (37)$\,\sim\,$(39) in Eqs. (35) and (36) the computation of the integrals of the approximate ones denoted by $A_0(p)$ and $B_0(p)$ 
is done making use of the dimensional regularization method and yields
\begin{eqnarray}
A_0(p) =\frac{12 M}{(4 \pi)^2\,\epsilon}\,(\frac{f_\pi}{m_\pi})^2\,
(M^2+m_\pi^2-\frac{M}{2}\,a-\frac{a^2}{2})+\,O(\epsilon^0),
\end{eqnarray}
\begin{eqnarray}
B_0(p) = 1-\frac{6}{(4 \pi)^2\,\epsilon}\,(\frac{f_\pi}{m_\pi})^2\,
(M^2+m_\pi^2-\frac{M}{2}\,a-\frac{a^2}{2})+\,O(\epsilon^0),
\end{eqnarray}
where $a \equiv \gamma\cdot p - M$ and the infinitesimal quantity $\epsilon$ comes from the shift of the dimension
($4\rightarrow4-\epsilon$) like that of the polarization function.
The calculation of $A_0(p)$ and $B_0(p)$ is seen in Appendix 2. 
\\\hspace*{4.mm}
The on-shell condition of the nucleon propagator is satisfied by adding the counter terms for the nucleon mass
and the wave function to the right side in Eq. (29) 
so as to subtract the first, the second and the third terms in Eq. (40).
As the consequence of it the renormalized self-energy $\Sigma^R (p)$ is obtained by $\epsilon\rightarrow 0$
as follows
\begin{eqnarray}
\Sigma^R(p) = \frac{M\,a^2}{M^2+m_\pi^2-\frac{M a}{2}-\frac{a^2}{2}}.
\end{eqnarray}
Unlike the usual calculation of this kind
the present result of the self-energy is not dependent on the coupling constant $f_\pi$.  
It attributes to the pseudovector coupling interaction.
Because of the derivative the estimate of the integral is $\Sigma_1 \sim k^3$ for the perturbative part of the self-energy 
and then the prescription by means of the counter terms does not eliminate all of the divergences
for the perturbative expansion.
In fact after the subtraction by the counter terms 
the remaining $\epsilon^{-1}$ dependent term in $A_0(p)$ cancels out that in $B_0(p)$ 
arising from the non-perturbative part of the vertex function. 
\hspace*{4.mm}
\section*{\normalsize{4 \quad Conclusion}}
\hspace*{4.mm}
According to the method in QED the relation between the nucleon propagator and the vertex part analogous to it
has been derived about the pseudovector coupling pion-nucleon system.
Since the axial-vector current is not conserved 
the relation acquires the non-perturbative term coming from the derivative on pion field 
in addition to the usual three-point vertex part determined by the perturbative expansion. 
As well as the pseudoscalar coupling the quantum corrections are found to be convergent also in the pseudovector coupling
within the normal way of removing the divergences by taking account of the non-perturbative term.
The application to the calculation of the propagators has given the unusual result particularly
on the self-energy of nucleon. 
The form of the propagator at the off-shell region 
may achieve the desired effect on the quantity relating to the higher-order process by pion 
along with the exact treatment of the polarization function.
\\\\\\\hspace*{4.mm}
{\large\bf Appendix 1}
\\\\\hspace*{4.mm}
The explicit form of $I(q)$ is
\\\begin{eqnarray}
I(q) \equiv \int_0^1 dz \, {\rm log} \, \frac{M^2-z(1-z)q^2}{4 \pi \mu^2} \qquad\qquad\nonumber
\end{eqnarray}
\begin{eqnarray}
=\, {\rm log} \, \frac{M^2}{4 \pi \mu^2} \,-2 
+2\,\sqrt{\frac{4M^2}{q^2}-1}\;{\rm arctan}\frac{1}{\sqrt{\frac{4M^2}{q^2}-1}}.
\end{eqnarray}\\
$I(q)$ is expanded in the series of $q^2-m_\pi^2$ and the coefficients $I_0$ and $I_1$ are given by
\begin{eqnarray}
I(q) = I_0 + I_1\,(q^2-m_\pi^2) + O((q^2-m_\pi^2)^2),
\end{eqnarray}
\begin{eqnarray}
I_0 ={\rm log} \, \frac{M^2}{4 \pi \mu^2} \,-2 +2\sqrt{\frac{4M^2}{m_\pi^2}-1}
\;{\rm arctan}\frac{1}{\sqrt{\frac{4M^2}{m_\pi^2}-1}},
\end{eqnarray}
\begin{eqnarray}
I_1 =\frac{1}{m_\pi^2}\,(1-\frac{4M^2}{m_\pi^2 \sqrt{\frac{4M^2}{m_\pi^2}-1}}
\;{\rm arctan}\frac{1}{\sqrt{\frac{4M^2}{m_\pi^2}-1}}).
\end{eqnarray}
\\\\\hspace*{4.mm}
{\large\bf Appendix 2}
\\\\\hspace*{4.mm}
$A_0(p)$ and $B_0(p)$ in Eqs. (40) and (41) are determined in terms of the quantity $J(p)$ as
\begin{eqnarray}
&&A_0(p) = 6\,i\,M\,(\frac{f_\pi}{m_\pi})^2\,\gamma_5\,J(p)\,\gamma_5, \\\nonumber\\
&&B_0(p) = 1-\frac{A_0(p)}{2 M},
\end{eqnarray}
\begin{eqnarray}
J(p) \equiv \int\frac{d^4 k}{(2 \pi)^4}
\,\frac{\gamma\cdot k\,(\gamma\cdot(p-k)+M)}{(k^2-m_\pi^2+i\epsilon)((p-k)^2-M^2+i\epsilon)}\nonumber
\end{eqnarray}
\begin{eqnarray}
=\frac{2\,i}{(4\pi)^2 \,\epsilon}(-M^2-m_\pi^2+\frac{M}{2}\gamma\cdot p+\frac{p^2}{2})+O(\epsilon^0).
\end{eqnarray}
\\\hspace*{4.mm}

\end{document}